\documentclass[12pt]{article}
\def \1{\'{\i}}
\def \n{\noindent}
\def \&{&=&}
\def \t{\paragraph{$\bullet$}}
\def \nn{\nonumber}
\newcommand{\be}{\begin{equation}}
\newcommand{\ee}{\end{equation}}
\newcommand{\beq}{\begin{eqnarray}}
\newcommand{\eeq}{\end{eqnarray}} 
\newcommand{\ba}{\begin{array}}
\newcommand{\ea}{\end{array}}  
\begin{document}
\setcounter{equation}{0}
\setcounter{section}{0}
\title{\Large \bf Miura Transformation between two Non-Linear Equations in $2+1$ dimensions}
\author{ \bf J.M. Cerver\'o  and P.G.
Est\'evez\footnote{e-mail: pilar@sonia.usal.es}  \\ {\small
\bf Area de F\1sica Te\'orica}\\ {\small \bf Facultad de F\1sica}\\ {\small \bf Universidad de
Salamanca}\\  {\small \bf 37008 Salamanca. Spain}\\} 
\maketitle
\begin{abstract} A Dispersive Wave Equation  in $2+1$ dimensions (2LDW) widely discussed by different
authors is shown to be nothing but the
modified version  of the Generalized Dispersive Wave Equation (GLDW). Using Singularity Analysis and
techniques based upon the Painlev\'e Property leading to the Double Singular Manifold Expansion we
shall find the Miura Transformation which converts the 2LDW Equation into the GLDW Equation. Through
this Miura Transformation we shall also present the Lax pair of the 2LDW Equation as well as some
interesting reductions to several already known integrable sytems in $1+1$ dimensions. As the 2LDW
Equation arises from a Miura Transformation we propose that it should be treated  conventionally as
a Modified Equation. In this case, we propose its designation as  {\bf The MGLDW Equation}  

\end{abstract} \vspace*{0.3in}

{\bf PACS Numbers 02.30 and 03.40K}
\newpage
\section{ Introduction}
\setcounter{equation}{0}

Recently the following system of non-linear partial differential equations in $2+1$ dimensions has
prompted a great deal of interest and several papers have appeared dealing with this system
\cite{Radha}, \cite{Porsezian}. It could also be considered as a Long Dispersive Wave Equation
(2LDW) in the form:
 
\beq \nonumber0&=&\lambda q_t+q_{xx}-2qV\\
 \nonumber 0&=&\lambda r_t-r_{xx}+2rV\\
 0&=&(qr)_x-V_y\Longrightarrow V=\int_{-\infty}^y(qr)_xdy' \eeq
  
\t For $\lambda=i$ and $r=q^*$, it is identical to the equation proposed by Fokas
\cite{Fokas} as the simplest integrable equation  in $2+1$ dimensions. In this reference, the author
presents the Lax pair of this particular case of (1.1). It is also worth pointing out that when 
$x=y$  the equation proposed by Fokas reduces to the Non-Linear Schr\"odinger Equation
\cite{Zakharov}
\t The real version of (1.1) has been obtained by Chakravarty, Kent and Newman \cite{Chakravarty}
as a particular reduction of the General Self-Dual Yang Mills Equations while the complex version
of (1.1) has recently been discussed by Maccari \cite{Maccari} using the so-called
Asymtotically Exact Reduction Method with the Kadomtsev-Petviashily Equation as a starting point.
\t The Painlev\'e Property \cite{WTC}, \cite{Weiss} as well as the Bilinear Hirota Equations for
the equation under consideration have been studied by Porsezian \cite{Porsezian} for the complex
version and by Radha and Lakhsmanan 
\cite{Radha} for the real case. In this latter reference, a systematic use of the Bilinear Formalism
is made in order to obtain soliton-like solutions of dromion type. 
 
\medskip

The aim of this paper is to show that equation (1.1) is nothing but the modified version of the
Generalized Dispersive Wave Equation (GLDW) which has been thoroughly studied by Boiti, Le\'on and
Pempinelli \cite{Boiti}. In order to show this, we shall be using  singularity
analysis of the GLDW that has already been succesfully applied  to this equation by one of the present
authors \cite{EG97}. Also using  techniques based upon Painlev\'e Analysis, we shall 
find the Miura transformation which converts (1.1) into the  GLDW. In view of these
results, it appears reasonable to refer to equation (1.1) as the Modified Generalized Dispersive Wave
Equation (MGDLW). Since the Lax pair of the GLDW equation is known \cite {AC}, \cite{Boiti}, the
Miura transformation that we shall present in this paper can easily be used to find the Lax pair
for   equation (1.1). As a final point, we shall offer several interesting reductions of our
equation  to some integrable equations in $1+1$ dimensions.

\section{ The Miura Transformation between the GDLW and the MGDLW}
\setcounter{equation}{0}
In 1987 Boiti {\it et al.} \cite{Boiti} proposed a dispersive wave equation in $2+1$ dimensions
\beq
\nonumber 0&=&u_{ty}+(\eta_{xy}+2uu_y)_x\\
0&=&\eta_{ty}+(u_{xy}+2u\eta_y)_x,\eeq which  we shall  hereafter call the Generalized Dispersive 
Long Wave Equation (GDLW) since  its reduction when $x=y$ is the dispersive wave equation also called
the Kaup System \cite{Kaup} or Classical Boussinesq System \cite{EG93}. Furthermore,  the
stationary version of (2.1) corresponds to the Sinh-Gordon Equation \cite {Boiti}, \cite{ECG}.

\paragraph{$\bullet$ The Painlev\'e Property}

One of the present authors \cite{EG97} has studied and discussed the system given by (2.1) from the
point of view of  Painlev\'e  Analysis.  As  is well known \cite{WTC} the Painlev\'e  Property
(PP), for practical purposes can be summarized  in the statement that all solutions of equation
(2.1) could be written in the form:

\beq \nonumber u=\sum_{j=0}^{\infty}u_j\chi^{j-a}\\
\eta=\sum_{j=0}^{\infty}\eta_j\chi^{j-b}\eeq

\noindent where $\chi=0$ is an arbitrary function called the Movable Singularities Manifold while
$u_j$ y $\eta_j$ are analytic functions in an arbitrary neighbourhood of this manifold. Inserting
(2.2) in (2.1), the dominant terms balance easily yields
\be a=b=1 \ee
\be u_o=\pm \chi_x\qquad\qquad \eta_o=\chi_x\ee

\noindent where $\pm$ means that two  Painlev\'e expansions are in principle possible.
In reference \cite{EG97} a detailed discussion is presented on how to deal with both expansions
simultaneously.

\paragraph{$\bullet$ The Double Singular Manifold Method}
In 1983 Weiss \cite{Weiss} proposed the Singular Manifold Method (SMM) which has been proved to
be extremely successful in dealing with a large class of non-linear equations. 
 The method is based on the 
restriction of the solutions of (2.1) for which  expansion (2.2) becomes truncated at the
constant level. In this case, one can prove that $\chi$ is no longer an arbitrary function
since it must verify the truncation condition. Accordingly, henceforth it is called {\it
Singular Manifold}. As  system (2.1) presents two different Painlev\'e branches, the
suggestion made in
\cite{EG97} is to use  {\bf two singular manifolds} $\phi$ and $\hat \phi$ (one for each expansion
branch) in such a way that the truncation of (2.2) adopts the form:
\beq \nonumber u'&=& u+{\phi_x\over \phi}-{\hat\phi_x\over \hat\phi}\\
\eta'&=& \eta+{\phi_x\over \phi}+{\hat\phi_x\over \hat\phi}\eeq

\paragraph{$\bullet$ Superposition of Solutions:} Expansion (2.5) suggests the following change
of dependent variables:  
\beq \nonumber u&=&m-\hat m\\
\eta&=&m+\hat m\eeq
such that $m$ and  $\hat m$  satisfy individual equations, each  with just one branch of the
Painlev\'e expansion. In order to find which equations $m$ y $\hat m$ should satisfy, we add and
substract (2.1), and using (2.6) we finally obtain:
\beq \nonumber 0&=& m_{ty}+\left[m_{xy}+2(m-\hat m)m_y\right]_x\\
0&=&\hat m_{ty}+\left[-\hat m_{xy}+2(m-\hat m)\hat m_y\right]_x\eeq

Let us now define
\beq \nonumber m_t&=&n_x\\ \hat m_t&=&\hat n_x\eeq
After an integration over the variable $x$ in (2.7) we obtain
\beq 0&=&n_y+m_{xy}+2(m-\hat m)m_y\\ 0&=&\hat n_y-\hat m_{xy}+2(m-\hat m)\hat m_y\eeq
These equations allow us to obtain  $m$ and $\hat m$ from (2.9) and (2.10). The result is:
\beq m&=& \hat m+{\hat m_{xy}-\hat n_y\over 2\hat m_y}\\ 
\hat m&=&  m+{ m_{xy}+ n_y\over 2 m_y}\eeq
By substituting  (2.11) in (2.9) and (2.12) in (2.10), one can easily decouple these equations, thus
obtaining one that is satisfied by $m$ and another one that is satisfied just by $\hat m$. Is is
trivial to check that both equations are one and the same but written with different variables.  In
fact
$(m,n)$ satisfies:
\beq \nonumber 0&=& m_t-n_x\\
\nonumber 0&=&m_y^2\left(n_{yt}-m_{xxxy}\right)+m_{xy}\left(n_y^2-m_{xy}^2\right)\\
&+&2m_y\left(m_{xy}m_{xxy}-n_yn_{xy}\right)-4m_y^3m_{xx}\eeq
while  $(\hat m,\hat n)$ satisfies:
\beq \nonumber 0&=& \hat m_t-\hat n_x\\
\nonumber
0&=&\hat m_y^2\left(\hat n_{yt}-\hat m_{xxxy}\right)+\hat
m_{xy}\left(\hat n_y^2-\hat m_{xy}^2\right)\\
&+&2\hat m_y\left(\hat m_{xy}\hat m_{xxy}-\hat n_y\hat n_{xy}\right)-4\hat
m_y^3\hat m_{xx}\eeq

\noindent which is obviously the same equation. Therefore, $m$ and $\hat m$ are both solutions of 
equation (2.13) but constrained by the the transformation  (2.11-12). It is not dificult to
recognize the constraint as an  {\bf Auto-B\"acklund Transformation} between  two solutions of
(2.13).

\paragraph{$\bullet$ The Miura Transformation:}With  regard to $u$ and $\eta$, these can be expressed
through (2.6) as a linear superposition of $m$ and $\hat m$. Taking into account (2.6) and (2.11),
one can write:
\beq \nonumber u&=&m-\hat m={\hat m_{xy}-\hat n_y\over 2\hat m_y}=-{ m_{xy}+ n_y\over
2 m_y}\\ \eta&=&m+\hat m =2\hat m+{\hat m_{xy}-\hat n_y\over 2\hat m_y}=2m+{ m_{xy}+
n_y\over 2 m_y}\eeq In order to express $m$ and $\hat m$ as a function of $u$ and $\eta$ with
the help of (2.15), we  must previously realize that the first of the equations (2.1) can be
integrated in the form:
\be \eta_x=-\partial_x^{-1}u_t-u_x^2\ee
in such a way that $m$ and $\hat m$ can be finally expressed as:
\beq \nonumber 2m_x&=&u_x+\eta_x=u_x-u^2-\partial_x^{-1}u_t\\
2\hat m_x&=&\eta_x-u_x=-u_x-u^2-\partial_x^{-1}u_t\eeq

\n Expressions (2.17) are Miura transformations between solutions of the non-linear systems
(2.1) and (2.14). In the next Section we shall show that  system  (2.14) is nothing but the
initial system given by (1.1). Obviously, (2.13) must be considered as the {\bf Modified GLDW
Equation} since it has been tranformed trough the {\bf Miura transformation} (2.17) from the GLDW
equation. Hereafter, we sahall refer to (2.13) as the MGLDW equation.

\section{The Modified GDLW Equation}
\setcounter{equation}{0}
We now attempt to show that (1.1) must truly be considered the modified GDLW equation
(2.13)
\subsection{Equation (1.1) as the Modified GDLW Equation}
We first try to write down (1.1) as a equation for just one field. In order to do so, we make the
change
\be V=-m_x\ee
in such a way that (1.1) becomes: 
\beq 0&=& q_t+q_{xx}+2qm_x\\
  0&=& r_t-r_{xx}-2rm_x\\
 0&=&qr+m_y \eeq
where  time has been rescaled in the form:
$$t\rightarrow \lambda t$$
Taking  $q$ out of  (3.4) and substituting it in (3.2), we find:
\be 0={m_{yt}+m_{xxy}\over r}-2m_{xy}{r_x\over r^2}+m_y\left({2m_x\over r}-{r_t\over
r^2}-{r_{xx}\over r^2}-2{r_x^2\over r^3}\right)\ee
 
\noindent Using (3.3) in (3.5), we also obtain
\be 0=m_{xxy}+m_{yt}-\left (2m_y{r_x\over r}\right)_x\ee
which can easily be integrated by setting $m_t=n_x$, which yields
 
\be {r_x\over r}={m_{xy}+n_y\over
2m_y}\ee Substituting (3.7) in (3.3), we obtain
\be{r_t\over r}=2m_x+{m_{xxy}+n_{xy}\over 2m_y}-{m_{xy}^2-n_y^2\over 4m_y^2}\ee
Next, we calculate the obvious identity  $\displaystyle{\left({r_t\over
r}\right)_x=\left({r_x\over r}\right)_t}$ using (3.7) and (3.8), and finally we obtain for $m$
the equation:
\beq \nonumber 0&=& m_t-n_x\\
\nonumber 0&=&m_y^2\left(n_{yt}-m_{xxxy}\right)+m_{xy}\left(n_y^2-m_{xy}^2\right)\\
&+&2m_y\left(m_{xy}m_{xxy}-n_yn_{xy}\right)-4m_y^3m_{xx}\eeq

\noindent which is precisely equation (2.13). Taking into account (3.1) and (2.17), we also find
\be V={1\over 2}\left (u_x+u^2+\partial_x^{-1}u_t\right)\ee
that establishes the relationship between the $V$-field of (1.1) and the $u$-field of (2.1).

Also, one can compare (3.7-8) and (2.15) yielding
\beq \nn {r_x\over r}\&-u\\{r_t\over r}\&\eta_x+u^2\eeq
\subsection{The Lax pair for MGDLW}
The Lax pair of the GLDW equation (2.1) is well known and can easily be written as \cite{EG97}:

\beq \nonumber 0&=&\varphi_t-\varphi_{xx}+2u\varphi_x
\\0&=&\varphi_{xy}-\partial_x^{-1}u_y\varphi_x+{\eta_y-u_y\over 2}\varphi
 \eeq

\noindent Alternatively, one can also write the Lax pair by using the gauge transformation: 

$$\varphi=e^{\partial_x^{-1}u}\psi $$
 
\n which leads equation (3.12) to the expression: 

\beq \nonumber
0&=&\psi_t-\psi_{xx}+\left[\partial_x^{-1}u_t-u_x+u^2\right]\psi
\\0&=&\psi_{xy}+u\psi_y+{\eta_y+u_y\over 2}\psi
 \eeq
This is the form of the Lax pair obtained by other authors  \cite{AC}, \cite{Boiti}.
Using (2.15-17), expression (3.13) becomes:

\beq \nonumber
0&=&\psi_t-\psi_{xx}-2 m_x\psi
\\0&=&\psi_{xy}-{ m_{xy}+ n_y\over 2 m_y}\psi_y+ m_y\psi,
 \eeq
which can be considered as the Lax pair for the MGLDW equation (2.13). 
Taking into account (3.1), (3.4) and (3.8), this latter expression can also be written as

\beq \nonumber
0&=&\psi_t-\psi_{xx}+2V\psi
\\0&=&\psi_{xy}-{ r_x\over r}\psi_y-qr\psi,
 \eeq
which would be the same Lax pair but for  form  (3.2-4) of the MGLDW equation. Since (3.1-4)
is invariant under the transformation
\beq\nonumber t&\rightarrow& -t\\
\nonumber r&\rightarrow& q,\eeq
the following pair of equations
\beq \nonumber
0&=&\hat \psi_t+\hat \psi_{xx}-2V\hat \psi
\\0&=&\hat\psi_{xy}-{ q_x\over q}\hat \psi_y-qr\hat \psi
 \eeq
would also be a Lax pair for the MGLDW equation.

\section{ Particular Cases}
\setcounter{equation}{0}
Let us now examine the Miura transformations induced in some interesting reductions of the
MGLDW equation:
\subsection{The stationary case: Sinh-Gordon versus AKNS} 
\t Let us assume that the fields of the {\bf MGDLW equation} (3.9) do not to depend on $t$. The
equation takes the form:
\be 0= -m_y^2m_{xxxy}-m_{xy}^3+2m_ym_{xy}m_{xxy}-4m_y^3m_{xx}\ee
which can be integrated with respect to $x$ to give:
\be 2m_{xxy}m_{y} +8m_{x}m_{y}^2-m_{xy}^2=0\ee
Eliminating $m_{xxy}$ between (4.1) and (4.2), we finally obtain:
\be m_{xxxy}+8m_{x}m_{xy}+4m_{y}m_{xx}=0,\ee
which is the {\bf AKNS equation in $1+1$ dimensions} \cite{AKNS}

\t With the same reduction of time-independent fields, {\bf The GLDW equation } (2.1) can also be
written as:
\beq  0&=& \eta_{xy} +2uu_y\\
0&=& u_{xy}+2u\eta_y\eeq
Let us now multiply (4.4) by $\eta_y$, (4.5) by $u_y$, and substract. The result is:  
$${d(\eta_y^2-u_y^2)\over dx}=0$$
Next, we introduce the change of variables
\beq \nn \eta_y&=&a_0\cosh 2q \\
u_y&=&a_0\sinh 2q\eeq
and so equations (4.4-5) take the form
\be
u= -q_x\ee 
Alternatively, using (4.6) we finally obtain:

\be q_{xy}=a_0\sinh 2q \ee
which is  {\bf  the Sinh-Gordon equation} \cite{AC}, \cite{ECG}

\t Therefore, the particularization of the Miura transformation (2.17) to the stationary case:
\beq \nn 2m_x&=&u_x-u^2=-q_{xx}-q_x^2\\
2\hat m_x&=&-u_x-u^2=q_{xx}-q_x^2\eeq
yields the corresponding Miura transformation between solutions of the AKNS equation (4.3) and
the Sinh-Gordon equation (4.8).
\subsection{Non-Local Boussinesq Equation versus the Kaup System}

In the next subsection we shall  deal with the reduction which arises from setting $x=y$

\t In the present case, the MGLDW equation (3.10) is reduced to:

\beq \nonumber 0&=& m_t-n_x\\
0&=&
m_x^2(n_{xt}-m_{xxxx})+m_{xx}(n_{x}^2-m_{xx}^2)\\
\nn \&2m_{x}(m_{xx}m_{xxx}-
n_{x}n_{xx})-4m_{x}^3m_{xx}\eeq
The set of equations (4.10) can be integrated with respect to $x$. One easily obtains:
\beq \nonumber 0\& m_t-n_x\\
0&=&
m_x(n_{t}-m_{xxx})-(n_{x}^2-m_{xx}^2)-2m_{x}^3\eeq

\n Alternatively, we can eliminate the field $n$. Thus, for $m$ we find  the following equation:

\be m_{tt}=\left[m_{xxx}+2m_x^2+{n_{x}^2-m_{xx}^2\over m_x}\right]_x\ee

\n Lambert et {\it al.} \cite{LW}, \cite{WL} have called (4.12) {\bf the non-local Boussinesq
equation (NLBq)}

\t We now apply the reduction $x=y$ to the GLDW equation (2.1):
\beq \nn 0\&u_t+\eta_{xx}+2uu_x\\
0\& \eta_t+u_{xx}+2u\eta_x\eeq
This system has been known as the classical Boussinesq system \cite{EG93} or {\bf the Kaup system}
\cite{CMP}, \cite{Kaup}, \cite{kw}, \cite{LW}.

\t The Miura transformation (2.17) establishes a relationship between the solutions of
The non-local Boussinesq equation (4.12) and the Kaup system (4.13):

\t An important subcase of this reduction is the {\bf  non-linear  Schr\"odinger equation}. To
see this, we must notice that the reduction $x=y$ applied to the MGLDW in its form given by
(3.2-4)   appears  as:

\beq \nn 0\& q_t+q_{xx}-2q^2r\\
0\& r_t-r_{xx}+2r^2q\eeq

\n which for imaginary time,  $t\rightarrow it$ and $q=r^*$, is clearly the non-linear Schr\"odinger
equation  \cite{Fokas}, \cite {Tabor}. 

\section{Conclusions}
\t Firs, starting from  Painlev\'e analysis of the Dispersive Long Wave Equation in $2+1$
dimensions (GDLW), we have been able to decompose the solutions of this equation as a {\bf
linear superposition} of two solutions of another equation in $2+1$ that has been called Modified
Dispersive Long Wave Equation (MGDLW).

\t The expansion of the solutions of the GDLW as a linear superposition of  its
modified version (MGDLW) is shown to be invertible and gives rise to the Miura transformation
between both equations. This Miura transformation has been  obtained explicitly.

\t In Section III we have shown that the MGLDW equation is nothing but the equation proposed by
several authors
\cite{Fokas}, \cite{Radha}, \cite{Maccari}, \cite {Porsezian} as one of the simplest possible
integrable equation in
$2+1$ dimensions. 

\t Since the Lax pair of the GDLW equation is already known \cite{Boiti}, the Miura transformation
allows one to pass from the GDLW equation to the MGDLW equation. In particular, we can easily find the
Lax pair of the latter equation. 

\t In Section IV some interesting reductions of the MGDLW equation are presented using the Miura
transformation between  GDLW and MGDLW. In this way, one can obtain the Miura transformations
between AKNS and Sinh-Gordon in $1+1$ and between non-local Boussinesq equation and the
Kaup system, also in $1+1$ dimensions. 

\t   In view of the above results one could conjecture that when two or more expansion branches in
the Painlev\'e expansion appear in a given non-linear partial differential equation, a  Miura
transformation exists that converts the initial equation into one with just one branch in this
expansion.

\t Systematic application of the Singular Manifold Method \cite{Weiss} and  use of it 
to find the Darboux transformation for the MGDLW equation pave the way for finding extremely
simple algorithms for the construction of solutions that generalize the ones already obtained 
by other authors \cite{Radha} using the bilinear formalism. We shall shortly be presenting these
results  in a forthcoming publication.

\paragraph{ACKNOWLEDGEMENTS} This research has been supported in part by the DGICYT  under contract
PB95-0947.

\end{document}